\documentclass[aps, prd,english,superscriptaddress,preprintnumbers,onecolumn,floatfix,nofootinbib]{revtex4-1}
\usepackage{amssymb}
\usepackage[T1]{fontenc}
\usepackage{latexsym,epsfig}
\usepackage[latin9]{inputenc}
\usepackage{graphicx}
\usepackage{amsfonts}
\usepackage{epsfig}
\usepackage{dcolumn}
\usepackage{bm}
\usepackage{babel}
\usepackage{hyperref}
\usepackage{amsmath,mathrsfs}

\makeatletter

\@ifundefined{textcolor}{}
{
 \definecolor{BLACK}{gray}{0}
 \definecolor{WHITE}{gray}{1}
 \definecolor{RED}{rgb}{1,0,0}
 \definecolor{GREEN}{rgb}{0,1,0}
 \definecolor{BLUE}{rgb}{0,0,1}
 \definecolor{CYAN}{cmyk}{1,0,0,0}
 \definecolor{MAGENTA}{cmyk}{0,1,0,0}
 \definecolor{YELLOW}{cmyk}{0,0,1,0}
 }

\makeatother

\begin{document}

\date{\today }
\title{Maximum Entropy Inferences on the Axion Mass in Models with
Axion-Neutrino Interaction}
\author{Alexandre Alves}
\affiliation{Departamento de F\'isica, Universidade Federal de S\~ao Paulo, UNIFESP, Diadema, Brasil}

\author{Alex G. Dias}
\affiliation{Universidade Federal do ABC, UFABC, \\
Centro de Ci\^encias Naturais e Humanas,  Santo Andr\'e - SP, 09210-170, Brasil} 

\author{Roberto da Silva}
\affiliation{Universidade Federal do Rio Grande do Sul, UFRGS, Instituto de F\'isica,  Porto Alegre - RS, 91501-970, Brasil}%
\email{roberto.silva@ufrgs.br}

\begin{abstract}
In this work we use the Maximum Entropy Principle (MEP) to infer the mass of
an axion which interacts to photons and neutrinos in an effective low energy
theory. The Shannon entropy function to be maximized is suitably defined in
terms of the axion branching ratios. We show that MEP strongly constrains
the axion mass taking into account the current experimental bounds on the
neutrinos masses. Assuming that the axion is massive enough to decay into
all the three neutrinos and that MEP fixes all the free parameters of the
model, %, \textbf{and taking into account a more restrict optimization}, 
the inferred axion mass is in the interval $0.1\ $eV$\ <m_{A}<0.2$ eV, which
can be tested by forthcoming experiments such as IAXO. However, even in the
case where MEP fixes just the axion mass and no other parameter, we found
that $0.1$ eV $< m_A < 6.3$ eV in the DFSZ model with right-handed
neutrinos. Moreover, a light axion, allowed to decay to photons and the
lightest neutrino only, is determined by MEP as a viable dark matter
candidate. 
%\textbf{Moreover, we show that having information about one of the
%parameters: (i) axion mass, (ii) neutrino mass, and (iii) ratio of the
%coupling constants, the other two parameters can be univocally determined by
%this principle}.
%Under hypothesis where the axion decays into the lightest
%neutrino, besides the photons, its mass could belongs to an interval which
%makes the axion a dark matter candidate. This last case restricts the
%lightest neutrino mass.
\end{abstract}

\maketitle

\section{Introduction}

Although the discovery of the Higgs boson and its properties have
represented a major advance for verifying the mass generation mechanism
through spontaneous symmetry breaking, along with its consequences, an
explanation for the values of most of the elementary particles masses is
still missing. It is understood in the Standard Model that the photon has
zero mass due to an unbroken gauge symmetry, and the weak vector bosons $W$
and $Z^{0}$ have interdependent masses resulting from the electroweak
symmetry breakdown. Still, according to the Standard Model, the Higgs boson
and all the charged fermions have arbitrary nonzero masses, with the
neutrinos being massless. This last feature is in contradiction with the
neutrinos oscillation phenomena, whose description requires nonzero neutrino
mass differences. As a matter of fact, the present experimental limits show
that neutrinos are ultralight compared to the other known massive particles.
All of this might suggests that a new principle or mechanism is necessary to
reach a more satisfactory understanding of the elementary particles masses.

It was found in Ref.~\cite{dEnterria:2012eip} that the peak of a function
constructed by multiplying the basic fourteen Standard Model branching
ratios of the Higgs boson decay channels occurs for a Higgs boson mass which
is in good agreement to the experimental value measured at the LHC.
Additionally, it was also mentioned in Ref.~\cite{dEnterria:2012eip} a
possible analogy with some sort of entropy arguing that the mass of the
Higgs boson has a value that allows for the largest number of ways of decays
into elementary particles. The work of Ref.~\cite{Alves:2014ksa} indeed
showed that the value of the Higgs boson mass results from the Maximum
Entropy Principle~(MEP) \cite{jaynes}, where the information entropy is
suitably defined in terms of the Higgs branching ratios into Standard Model
particles, furnishing the most accurate Higgs boson mass theoretical
determination to date.

Our premise in invoking MEP is to consider an ensemble of $N$
non-interacting identical spinless particles which have $m$ basic decay
modes. The probability that the ensemble evolve to a final state
configuration, with $n_{0}$ bosons decaying into the mode with branching
ratio $BR_{0}$, $n_{1}$ bosons decaying into the mode with branching ratio $%
BR_{1}$, and so on until $n_{m}$ bosons decaying into the mode having
branching ratio $BR_{m}$, is given by the following multinomial distribution 
\begin{equation}
\mathrm{Pr}(\{n_{k}\}_{k=0}^{m})\equiv \frac{N!}{n_{0}!n_{1}!\cdots n_{m}!}%
\prod_{k=0}^{m}(BR_{k})^{n_{k}},  \label{Probability_distribution}
\end{equation}%
in which $\sum_{k=0}^{m}BR_{k}=1$, and $\sum_{k=0}^{m}n_{k}=N$. From these
probabilities, the Shannon entropy~\cite{shannon} associated to the
evolution of the initial ensemble to the final state in which all $N$
scalars have decayed is given by 
\begin{equation}
S_{N}=-\sum_{\{n_{i}\}}^{N}\mathrm{Pr}(\{n_{k}\}_{k=0}^{m})\,\mathrm{ln\,Pr}%
(\{n_{k}\}_{k=0}^{m})\,,  \label{sentropy}
\end{equation}%
with $\sum_{\{n_{i}\}}^{N}(\bullet )\equiv
\sum_{n_{0}=0}^{N}\sum_{n_{1}=0}^{N}\cdots \sum_{n_{m}=0}^{N}(\bullet
)\times \delta \left( N-\sum_{i=0}^{m}n_{i}\right) $~\cite{Alves:2014ksa}.
The entropy $S_{N}$ is a function of unknown quantities as masses of
particles and coupling constants entering in the branching ratios $BR_{k}$.
We propose that such quantities can be inferred through maximization of $%
S_{N}$, also taking into account constraints that may enter as prior
information. Similar approaches using information and configurational
entropies can be found, for example, in Refs.~\cite{mep}.

Our aim in this work is to use MEP to infer the mass of the axion, taking
into account a model which the axion decays to neutrinos and photons. More
specifically, we made the following assumptions on the axion field, $A(x)$.
First, its particle excitation, the axion, decays dominantly into a pair
photons and also into a pair of neutrinos. Second, its low energy effective
Lagrangian describing the interactions with photons and neutrinos is given
by 
\begin{equation}
\mathscr{L}_{eff}=\frac{1}{2}\partial _{\mu }A\,\partial ^{\mu }A-\frac{1}{2}%
m_{A}^{2}\,A^{2}-\frac{g_{A\gamma }}{4}\,A\,F_{\mu \nu }{\widetilde{F}}^{\mu
\nu }-\frac{g_{A\nu }}{2}\,\overline{\nu }_{i}\gamma ^{\mu }\gamma _{5}\nu
_{i}\,\partial _{\mu }A  \label{axlag}
\end{equation}%
where $F^{\mu \nu }$ is the electromagnetic field strength with $\widetilde{F%
}^{\mu \nu }\equiv \epsilon ^{\mu \nu \lambda \rho }F_{\lambda \rho }/2$ its
dual; $\nu _{i}(x)$, $i=1,\,2,\,3$, denotes the mass eigenstates neutrinos
fields; $m_{A}$ is the axion mass; $f_{A}$ the axion decay constant, which
is a high energy scale; $g_{A\gamma }$ and $g_{A\nu }$ are the axion-photon
and axion-neutrino coupling constants, respectively. Another important
remark is that we are tacitly assuming that neutrinos are Dirac fermions.
Therefore, the coupling between the axion and the neutrinos vanish in the
massless limit. In Appendix \ref{app1} we show an example of an ultraviolet
completed model leading to the effective Lagrangian in Eq. (\ref{axlag}).

The axion is a hypothetical pseudo Nambu-Goldstone boson remnant of an
anomalous $U(1)$ symmetry, spontaneously broken at the energy scale $f_{A}$,
present in extensions of the Standard Model motivated to solve the the
strong CP problem through the Peccei-Quinn mechanism~\cite%
{Peccei:1977hh,Weinberg:1977ma,Wilczek:1977pj} (for a review of the strong
CP problem and axions see, for example, Refs.~\cite%
{Kim:2008hd,Jaeckel:2010ni}). The actual constraints from the experiments
searching for the axion limits $f_{A}$ to be much above the electroweak
scale, i. e., $f_{A}\gg v=246$ GeV. Consequently, the axion interacts very
weakly with all other particles by the reason that the associated coupling
constants are suppressed by $f_{A}$. We define the axion-neutrino coupling
constant in Eq. (\ref{axlag}) as 
\begin{equation}
g_{A\nu }=\frac{C_{A\nu }}{f_{A}},
\end{equation}%
where $C_{A\nu }$ is the coefficient of the axion-neutrino coupling which
depends on the ratios of vacuum expectation values (see Appendix \ref{app1}%
). The axion-photon coupling constant is, by its turn, 
\begin{equation}
g_{A\gamma }=\frac{\alpha }{2\pi f_{A}}{\widetilde{C}}_{A\gamma }\,,
\label{gamax}
\end{equation}%
with $\alpha $ the fine structure constant, and the coefficient of the
axion-photon coupling 
\begin{equation}
{\widetilde{C}}_{A\gamma }=\left( \frac{C_{a^{\prime }\gamma }}{C_{a^{\prime
}g}}-\frac{2}{3}\,\frac{4+m_{u}/m_{d}}{1+m_{u}/m_{d}}\right) \,.
\label{ctil}
\end{equation}%
In the coefficient ${\widetilde{C}}_{A\gamma }$ the anomaly coefficients $%
C_{a^{\prime }\gamma }$ and $C_{a^{\prime }g}$ are model dependent and
typically of order one, with $m_{u}/m_{d}\approx 0.56$ the ratio of up and
down quark masses. Such a coefficients for different models, as well as
other features of the axion, can be found in~\cite{Dias:2014osa}.
Additionally, the axion mass is also suppressed by the energy scale $f_{A}$
and given by~\cite{Weinberg:1977ma} 
\begin{equation}
m_{A}\simeq \frac{m_{\pi }f_{\pi }}{f_{A}}\frac{\sqrt{m_{u}/m_{d}}}{%
1+m_{u}/m_{d}}\approx 0.48\frac{m_{\pi }f_{\pi }}{f_{A}},  \label{mafa}
\end{equation}%
in which $m_{\pi }$ and $f_{\pi }$ are the pion mass and its decay constant,
respectively. This makes the axion an ultralight particle for $f_{A}\gg 246$
GeV.

Our paper is organized as follows: In the next section we briefly discuss as
we intend to use the MEP in order to obtain the axion mass considering the
lightest neutrino mass and the coupling constants as inputs. In section \ref%
{Sec:Results_I} we present a first part of our results which consider axions
decaying into pairs of neutrinos and pair of photons, and a second part of
the results corresponding to axions decaying only into pair of the lightest
neutrinos, in addition to a pair of photons, which is presented in \ref%
{Sec:Results_II}. We present our conclusions in section \ref{Sec:Conclusions}

\section{Parameters inferences from MEP}

\label{Sec:Parameters_inference_from_mep}

Let us consider an initial state ensemble with a very large number $N$ of
axions. After a time $t\gg 1/\Gamma $, with the total axion decay width
given by the sum of the partial decay widths into a pair of photons, $\Gamma
_{0}$, plus the ones into pairs of neutrinos, $\Gamma _{i}$, i. e., 
\begin{equation}
\Gamma =\Gamma _{0}+\sum_{i}\Gamma _{i}\;,  \label{gammas}
\end{equation}%
the initial state ensemble evolves to a final state bath of photons and
neutrinos. The summation above (and below) extends only over those neutrinos
whose masses are less than $m_{A}/2$. If the axion is massive enough it
could decay in all three active neutrinos, where $i=1,\,2,\,3$.

The axion partial decay widths derived from the effective Lagrangian in Eq. (%
\ref{axlag}) are 
\begin{eqnarray}
\Gamma _{0} &=&\frac{g_{A\gamma }^{2}}{64\pi }m_{A}^{3}\,, \\
\Gamma _{i} &=&\frac{g_{A\nu }^{2}}{8\pi }m_{A}m_{{i}}^{2}\beta _{i}\,,
\label{adw}
\end{eqnarray}%
where $m_{i}$ is the $i$-th neutrino mass and $\beta _{i}=\sqrt{1-\frac{%
4m_{i}^{2}}{m_{A}^{2}}}$.

These widths lead to the following branching ratios for the axion decaying
into a pair of photons and into pairs of neutrinos 
\begin{eqnarray}
BR_{0} &=&\frac{\Gamma _{0}}{\Gamma }=\left[ 1+\sum_{i}32\pi ^{2}\frac{%
r_{\nu }^{2}}{\alpha ^{2}}\frac{m_{i}^{2}\beta _{i}}{m_{A}^{2}}\right]
^{-1}\,,  \label{b0} \\
BR_{i} &=&\frac{\Gamma _{i}}{\Gamma }=\left[ 1+\frac{\alpha ^{2}}{32\pi
^{2}r_{\nu }^{2}}\frac{m_{A}^{2}}{m_{i}^{2}\beta _{i}}+\sum_{j\neq i}\frac{%
m_{j}^{2}\beta _{j}}{m_{{i}}^{2}\beta _{i}}\right] ^{-1}\,,  \label{bi}
\end{eqnarray}%
in which we define $r_{\nu }=\left\vert C_{A\nu }/\widetilde{C}_{a\gamma
}\right\vert $ as the ratio of the anomaly coefficients of the
axion-neutrino coupling, ${C}_{A\nu }$, and the axion-photon coupling, ${%
\widetilde{C}}_{A\gamma }$. This ratio is equivalent to the ratio of the
associated coupling constants given by $\left\vert g_{A\nu }/g_{A\gamma
}\right\vert =2\pi r_{\nu }/\alpha $.

There are many possible final states characterized by the number of axions
which decay into a pair of photons, $n_{0}$, and by the numbers axions which
decay into each possible pair of neutrinos, $n_{i},\;i=1,2,3$. Considering
that the axion can decay into the three neutrinos plus the photon, according
to Eq.~(\ref{Probability_distribution}) the probability that the ensemble of 
$N$ axions decay into a particular final state characterized by $n_{0}$, $%
n_{1}$, $n_{2}$, and $n_{3}$ is 
\begin{equation}
\mathrm{Pr}(\{n_{k}\}_{k=0}^{3})=\frac{N!}{n_{0}!n_{1}!n_{2}!n_{3}!}%
\prod_{k=0}^{3}(BR_{k})^{n_{k}}\;.
\end{equation}%
The entropy function is then constructed from Eq.~(\ref{sentropy}) summing
over all the partitions satisfying $\sum_{k=0}^{3}n_{k}=N$.

According to MEP, the initial ensemble evolves to a final state of maximum
entropy permitting us to infer the values of the unknown quantities through
maximization of $S_{N}$. We do this taking into account the prior
information of the neutrinos mass from the best fit mass squared differences
determined by the data of neutrinos oscillations, considering the normal
hierarchy pattern~\cite{King:2017}, 
\begin{equation}
\begin{array}{ccccc}
\Delta m_{12}^{2} & = & m_{2}^{2}-m_{1}^{2} & = & (7.45\pm 0.25)\times
10^{-5}\,\mathrm{eV}^{2} \\ 
&  &  &  &  \\ 
\Delta m_{31}^{2} & = & m_{3}^{2}-m_{1}^{2} & = & (2.55\pm 0.05)\times
10^{-3}\,\mathrm{eV}^{2}\text{.}%
\end{array}
\label{Eq:difference_mass}
\end{equation}%
Thus, $S_{N}$ depends, effectively, on the three parameters $m_{A}$, $r_{\nu
}$, and the lightest neutrino mass which we denote as $m_{\nu }$. Finally,
our analysis do not depend on the neutrinos mass hierarchy pattern, so that
the same results are obtained if the inverted hierarchy is assumed.

Contrary to the situation of the Higgs boson mass inference~done in ref. 
\cite{Alves:2014ksa}, where the Higgs mass was the last independent Standard
Model parameter to be determined, in the model under study we have three
independent parameters as we just discussed. In principle, MEP could force
all these parameters to be fixed at the global maximum of the entropy.
However, this is not the case of the Standard Model, for example. The Higgs
mass does not correspond to a global maximum of $S$, but just to a
constrained maximum. Some parameters of the Standard Model are related by
its symmetries imposing strong constraints on these parameters, for example,
the $W$ and $Z$ bosons masses. This fact reveals that not all the parameters
of the model might be determined by the MEP. Nevertheless, the success in
the Higgs mass prediction suggests that MEP can be useful in inferring the
masses of scalar particles when the other parameters are fixed by some
different mechanisms.

In this work we remain agnostic about the type of inference that MEP can
actually perform, waiting for the experimental evidence to settle that.
Therefore, we take two main hypothesis: first, the entropy function is
maximized in $m_{A}$ only, with the other parameters considered as prior
information, that is, $S_{N}\equiv S_{N}(m_{A}|m_{\nu },r_{\nu })$; second,
a more restrict hypothesis: $S_{N}\equiv S_{N}(m_{A},m_{\nu },r_{\nu })\ $
where we will show that MEP can be more determinant in the sense that having
information about one of the parameters: (i) axion mass, (ii) neutrino mass,
and (iii) ratio of the coupling constants, the other two parameters can be
univocally determined by this principle.

The hypothesis in which $S_{N}\equiv S_{N}(m_{A}|m_{\nu },r_{\nu })$ has a
more direct analogy with thermodynamics where the entropy is a function of
the energy, which in our case is the axion mass. The ratio $r_{\nu }$\ and
the lightest neutrino mass $m_{\nu }$\ are free parameters which should be
determined prior to the inference of axion mass. However it is important to
mention that Shannon entropy in our formulation is not defined as a entropy
per particle, as function of an energy per particle, as is usual in
Thermodynamics. This does not spoil our study and deserves a deeper
discussion relative to our inference process. First of all, our weights in
Shannon entropy are not Boltzmann weights of some known problem of
interacting particles in contact with a thermal reservatory exactly as is
studied in Statistical Mechanics. Only in this situation, the Shannon
entropy should be equivalent to Boltzmann entropy definition. Nevertheless,
even without this equivalence it is important to mention that MEP is a
general and universal method than Statistical Mechanics approach, and its
basis are governed by probability theory. In this context we can write the
entropy of any probability distribution and maximize it in relation to their
physical parameters.

We impose the most recent experimental constraints from the neutrino
oscillation experiments summarized in Ref.~\cite{King:2017} 
\begin{equation}
m_{1}=m_{\nu }\;\;,\;\;m_{2}=\sqrt{m_{\nu }^{2}+7.45\times 10^{-5}}%
\;\;,\;\;m_{3}=\sqrt{m_{\nu }^{2}+2.55\times 10^{-3}}\;.
\label{vinculo-massas}
\end{equation}

The upper bound from the Planck Collaboration measurements of CMB
anisotropies~\cite{Ade:2015xua} for the sum of the neutrino masses, along
with Eq.~(\ref{vinculo-massas}) for a massless lightest neutrino, translate
to the following constraint that will be taken into account in our inference
process 
\begin{equation}
0.059\text{ eV}<\sum_{i=1}^{3}m_{i}<0.23\;\text{eV,}  \label{planck}
\end{equation}%
which implies the interval $0<m_{\nu }<0.0712$ eV, for the lighest neutrino
mass. In the next sections we will present our results.

\section{Results I: Axions decaying into the three neutrinos and photons}

\label{Sec:Results_I}

First, we suppose that the axion is heavy enough to decay into a pair of
photons and all the three neutrinos.

\begin{figure}[t]
\begin{center}
\includegraphics[width=0.65\columnwidth]{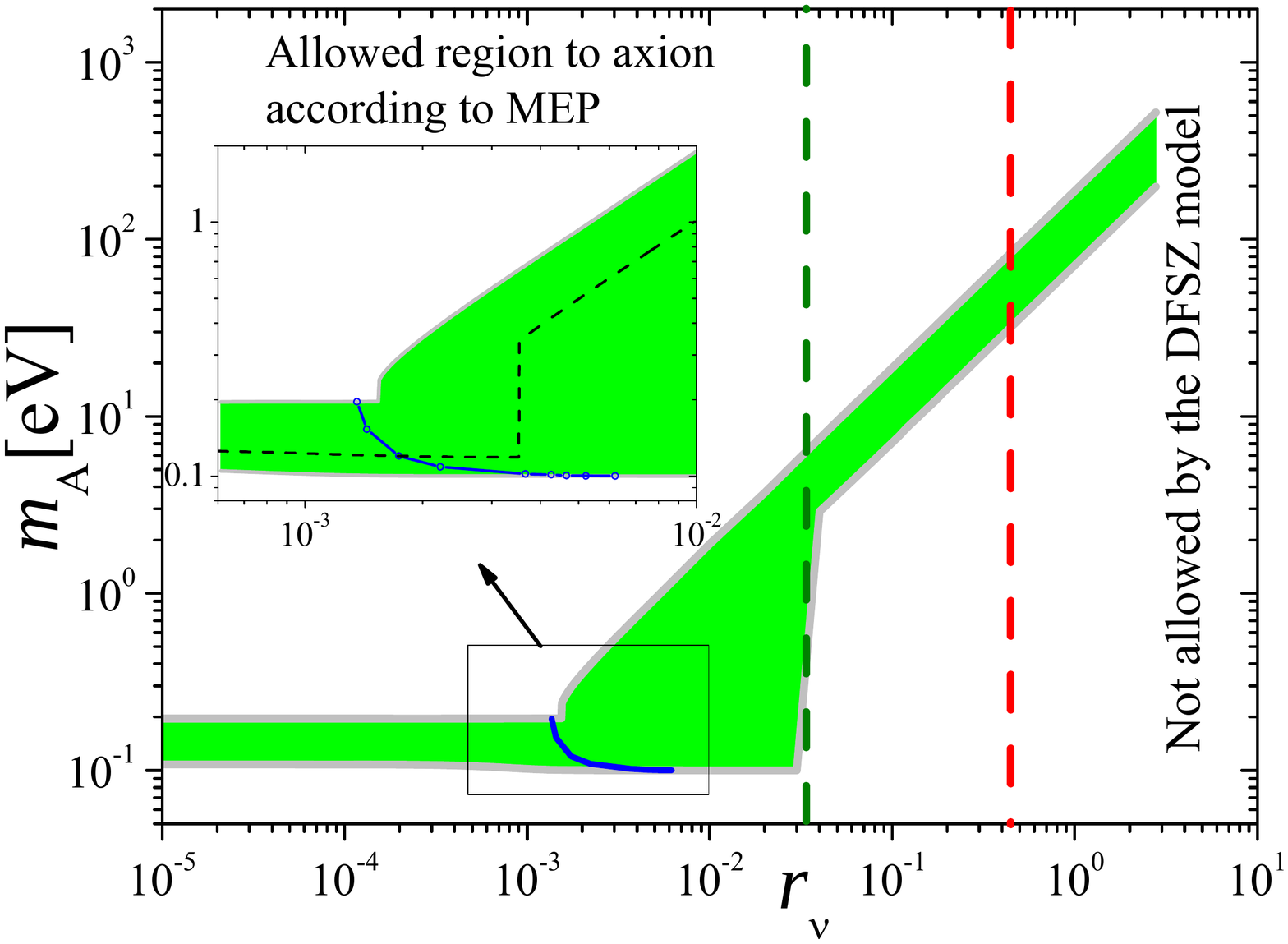}
\end{center}
\caption{Inferred axion mass region from MEP, taking into account the
constraints from Eq.~(\protect\ref{planck}). The gray curves correspond to
axion mass maximizing the entropy for the case in which the maximum lightest
neutrino mass (upper curve) and the minimum lightest neutrino mass (lower
curve) are assumed. The light green shaded region correspond to all maximal
entropy points, assuming that the lightest neutrino mass is in the interval
allowed by Eq. \protect\ref{planck}. The blue curve assumes that the maximum
of entropy fixes all the three parameters as discussed in text. The inset
plot corresponds to a zoom of the blue line, where the dashed curve is only
an example of curve in the allowed region corresponding to a intermediate
value of lightest neutrino mass: $m_{\protect\nu }=0.03$ eV between the
bounds. This curve cross the blue line in the point $(r_{\protect\nu %
},m_{A})\equiv (0.00174,0.12)$ which is the point of the highest entropy in
this dashed line. Finally the dashed red line shows the limit to acceptable
values of $r_{\protect\nu }$ for the DFSZ-type model which we present in
Appendix \protect\ref{app1} while the dashed dark green line shows the limit
from astrophysics for the $r_{\protect\nu }$ (see the text). }
\label{maximal}
\end{figure}

The entropy, given by Eq.~(\ref{sentropy}), can be written as \cite{Cichon} 
\begin{equation}
S_{N}(m_{A}|m_{\nu },r_{\nu })=S(m_{A}|m_{\nu },r_{\nu })+\ln (2\pi N)+O(%
\frac{1}{N^{2}})
\end{equation}%
with 
\begin{equation}
S(m_{A}|m_{\nu },r_{\nu })=\ln (BR_{0}BR_{1}BR_{2}BR_{3})  \label{Entropy}
\end{equation}%
where the branching ratios are related as $\sum_{i=0}^{3}BR_{i}=1$.

\begin{figure}[t]
\begin{center}
\includegraphics[width=0.5\columnwidth]{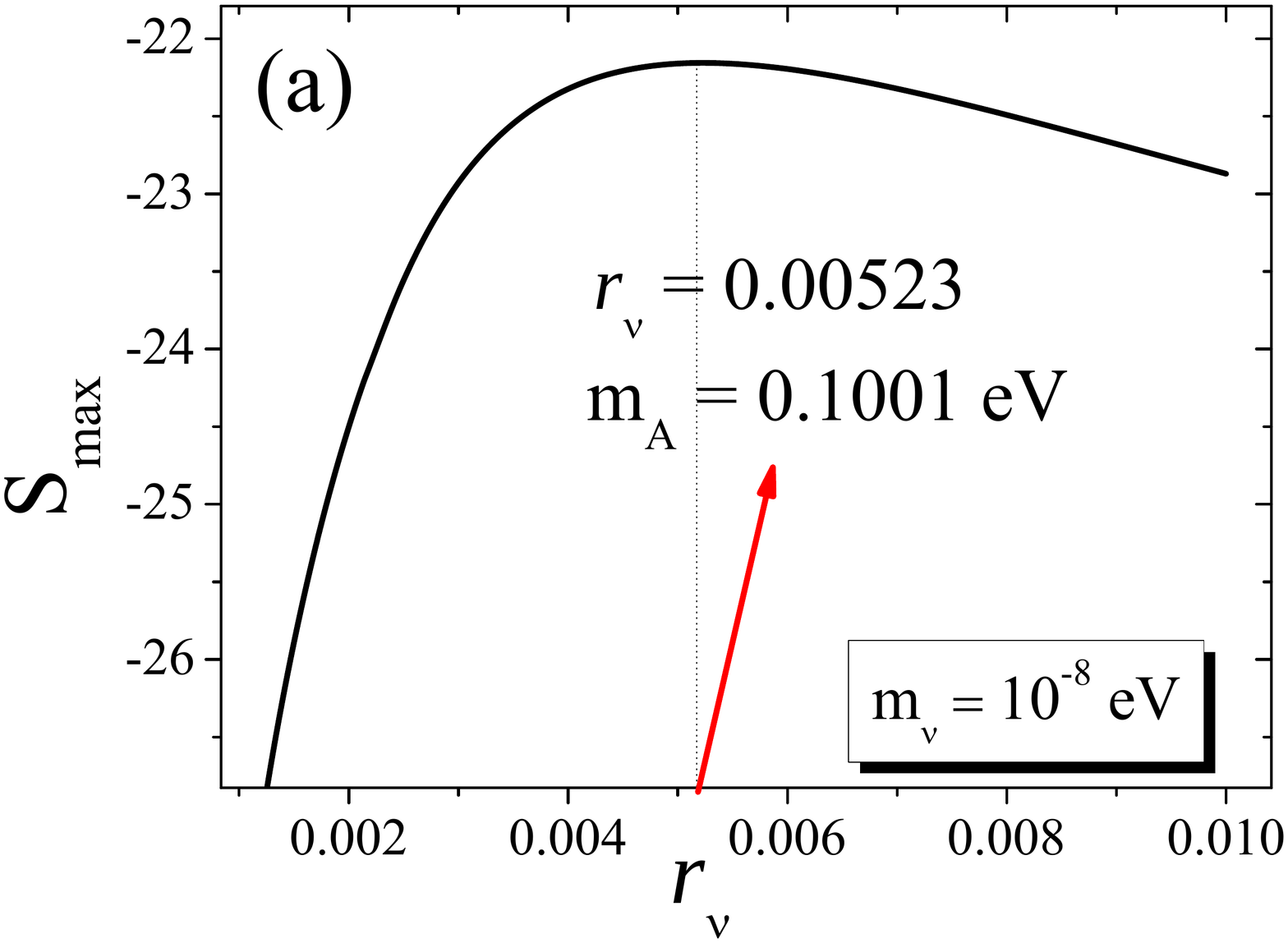}%
\includegraphics[width=0.5\columnwidth]{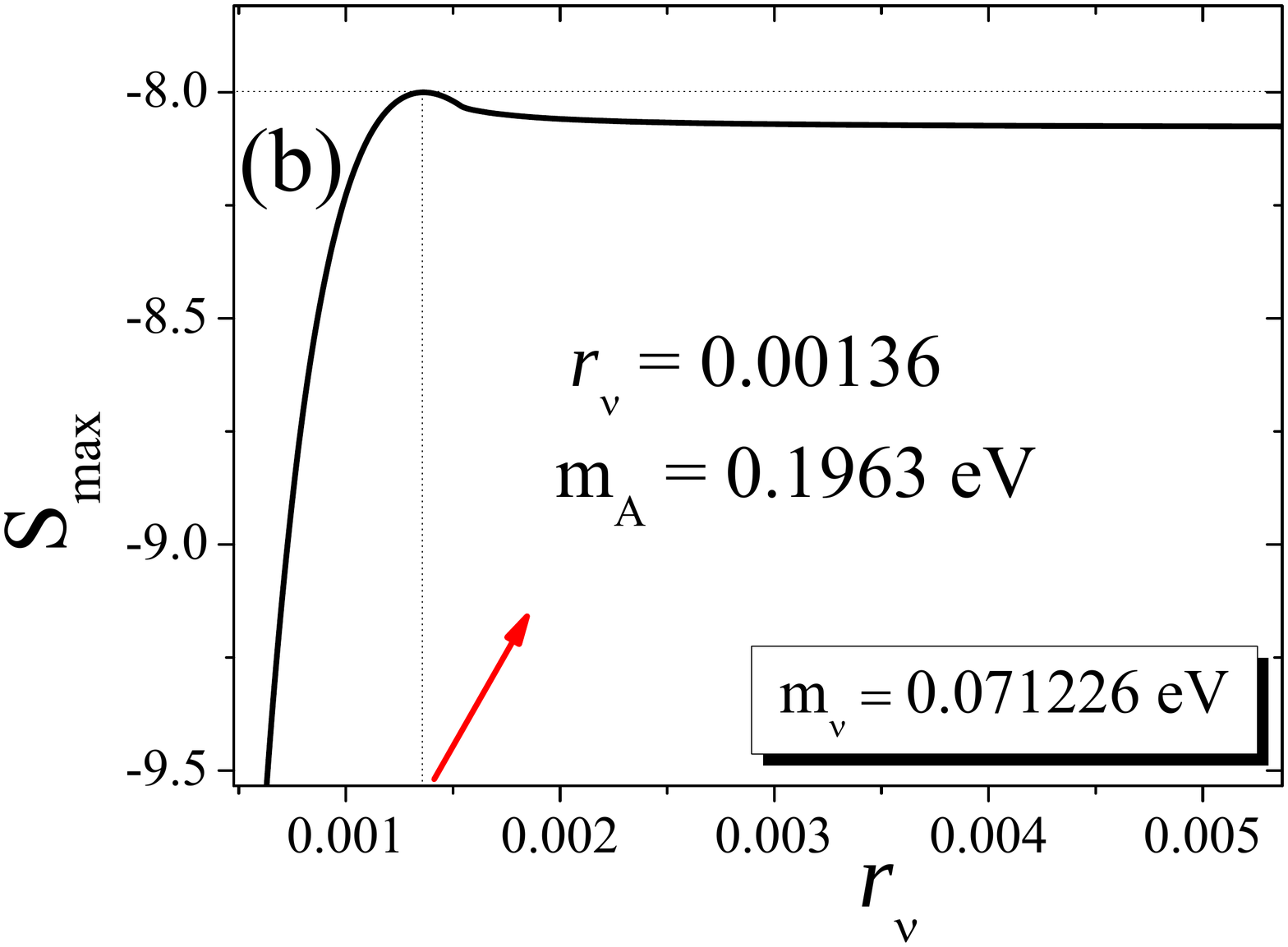}
\end{center}
\caption{The following plots (a) and (b) show the maximum entropy values
assumed in each point of the gray lines (see Fig.\protect\ref{maximal})
respectively for minimum ($m_{\protect\nu }=10^{-8}$ eV) and maximum ($m_{%
\protect\nu }=0.0721226$ eV) lightest neutrino mass. Both gray lines cross
the blue line (see again Fig. \protect\ref{maximal}) exactly in the points
that correspond to peak presented in the plots (a) and (b). }
\label{maximal_II}
\end{figure}

It can be shown that the global maximum of $S$ is obtained when all the
branching ratios are equal 
\begin{equation}
BR_{i}(m_{A},m_{\nu },r_{\nu })=\frac{1}{4}\;,\;i=0,1,2,3  \label{brssol}
\end{equation}%
However, it is also possible that this type of solution cannot be attainable
if further constraints arise from an UV complete theory in which the
parameters do not allow that the Eq. (\ref{brssol}) be satisfied. If further
constraints are absent or if they are weak, then we should expect that
nearly equal branching ratios constitute another prediction coming from MEP.

First of all we investigate the entropy supposing that MEP fixes just the
axion mass, i.e., $S=S(m_{A}|m_{\nu },r_{\nu })$, with the other parameters
either previously known or at least bounded by some data. As we have just
discussed, this might be the case if the UV complete theory fixes somehow
the $r_{\nu }$\ of the model and possibly other parameters. In our approach,
we, therefore, allow $m_{\nu }$\ to vary according to the constraints of
Eqs.~(\ref{vinculo-massas}), (\ref{planck}). In this case, for each fixed $%
m_{\nu }$ and $r_{\nu }$, we just seek for a solution ($m_{A}$) that
maximizes $S(m_{A}|m_{\nu },r_{\nu })$\ given by Eq.~(\ref{Entropy}).\ The
light green shaded area of Figure~\ref{maximal} shows the MEP inference for
the axion mass for the range of the $r_{\nu }$\ taking into account the
neutrino masses constraints from Eq.~(\ref{planck}).

The gray curves in Figure \ref{maximal} correspond to axion mass maximizing
the entropy for the case in which the maximum lightest neutrino mass (upper
curve) and the minimum lightest neutrino mass (lower curve) were assumed.
Such curves can be identified as iso-lightest-neutrino-mass curves in the
diagram $r_{\nu }\times m_{A}$. The following plots (a) and (b), in Figure %
\ref{maximal_II} show the maximum entropy values assumed in each point of
gray lines respectively for minimum ($m_{\nu }=10^{-8}$\ eV) and maximum ($%
m_{\nu }=0.0721226$\ eV) lightest neutrino mass. It is important to mention
that even when we used $m_{\nu }=10^{-7}$\ eV the same result was obtained
and the gray line remains the same. Obviously, numerically for $m_{\nu }=0$\
eV we have no results.

This suggests that we can look to MEP in more restrict optimization than
just seeking for a solution $m_{A}$, given $m_{\nu }$ and $r_{\nu }$. Then a
stronger inference can be made optimizing $S\mathbf{\equiv }S(m_{A},m_{\nu
},r_{\nu })$ which result is represented by the blue line where given one of
the parameters, is possible to obtain the other two ones. We give a more
detailed explanation as follows. The boundary gray lines cross the blue line
exactly in the points that correspond to peak presented in the plots (a) and
(b) in the same Figure \ref{maximal_II}. The inset plot in Figure \ref%
{maximal} corresponds to a focused area of the blue line region. Just to
give an example for $r_{\nu }=0.00174$, the point over the blue line, of the
highest entropy in the green region, corresponds to $m_{A}=0.12$ eV and $%
m_{\nu }=0.03$ eV. In this same inset plot, the dashed black curve in the
allowed region corresponds exactly to the intermediate value of lightest
neutrino mass: $m_{\nu }=0.03$ eV. This iso-lightest-neutrino-mass curve is
exactly between the bounds lines (gray lines) as expected. Therefore, the
light green region can be understood as a family of
iso-lightest-neutrino-mass curves of the diagram $r_{\nu }\times m_{A}$,
which cross the blue line which is composed by the optimal values found in
each iso-lightest-neutrino-masses curves. In other words, the blue line is
composed by values which MEP fixes all the three parameters.

The dashed red line in Figure \ref{maximal} shows the limit to allowed
values of $r_{\nu }$ for the DFSZ-type model with right-handed neutrinos
which we present in Appendix \ref{app1}. In the same Figure the dashed dark
green line shows the limit from astrophysics on the coupling (see below).

The inference is stronger for small $r_{\nu }$\ up to $\sim 0.0017$, for
higher values, the upper boundary of the Figure~\ref{maximal} increases
towards higher axion masses while the lower boundary remains nearly constant
with $r_{\nu }$. Note that, under this hypothesis, we need to know $r_{\nu }$%
\ and $m_{\nu }$\ in order to get $m_{A}$. The DFSZ model with right-handed
neutrinos restricts $r_{\nu }<0.46\cos ^{2}\beta $, which determines the
dashed red line (see Appendix \ref{app1}).

We observe that under the consideration of more restrict optimization (blue
line in Fig. \ref{maximal}), MEP constrains the axion mass to lie within 
\begin{equation}
0.1\;\text{eV}<m_{A}<0.2\;\text{eV}.  \label{inference_mA}
\end{equation}%
We observe that the interval in Eq. (\ref{inference_mA}) is in the threshold
of the projected sensitivity of the IAXO experiment~\cite{Armengaud:2014gea}%
. Thus, our hypothesis can tested experimentally in the near future.

We should mention that the interval of Eq. (\ref{inference_mA}) for the
axion mass derived from MEP is compatible with the limits from astrophysics
(compilation of the actual astrophysical limits on the axion mass and
coupling constants are given in~Ref. \cite{pdg}). For example, studies
concerning the evolution of stars on the horizontal branching~\cite%
{Ayala:2014pea} put the constraint $m_{A}<0.5$ eV ($f_{A}>1.3\times 10^{7}$
GeV) on the DFSZ model. There is an astrophysical limit that could
potentially impact on the interval in Eq. (\ref{inference_mA}), but it also
depends on the axion-electron coupling. A bound from red giants in the
Galactic globular cluster M5 provided $m_{A}\,\cos ^{2}\beta <15.3$ meV at $%
95\%$ CL~\cite{Viaux:2013lha}, in the DFSZ model having the axion-neutrino
coupling coefficient $C_{Ae}=\cos ^{2}\beta /3$\footnote{%
In the DFSZ model this happens to be the case in which the right-handed
electron field couples to the same Higgs doublet that give mass to the
u-type quarks. If, on the other hand the right-handed electron field couples
to the same Higgs doublet that give mass to the d-type quarks then the
axion-neutrino coupling coefficient turns out to be $C_{Ae}=-\mathrm{sin^{2}}%
\beta /3$, as can be seen in Appendix \ref{app1}, and the corresponding
astrophysical limits turns out to be on $m_{A}\,{\mathrm{sin^{2}}\beta }$.}.
Taking into account Eq. (\ref{inference_mA}) it means that $\cos ^{2}\beta
<0.0752$ implying that $r_{\nu }<0.034$. This restriction on $r_{\nu }$ is
shown in Figure~\ref{maximal} (dark green dashed line) which leads to $0.1$%
~eV~$< m_{A}< 6.3\ $ eV for the DFSZ model with the right-handed neutrinos
which we present in Appendix \ref{app1}.

\section{Results II: Axions decaying into the lightest neutrinos and photons.%
}

\label{Sec:Results_II}

Let us suppose now that the axion has only two decay modes: one into a pair
of photons, and other into a pair of the lightest neutrinos. In this case,
the two relevant branching ratios are 
\begin{equation}
BR_{0}=\left[ 1+32\pi ^{2}\frac{r_{\nu }^{2}}{\alpha ^{2}}\frac{m_{\nu }^{2}%
}{m_{A}^{2}}\left( 1-\frac{4m_{\nu }^{2}}{m_{A}^{2}}\right) ^{1/2}\right]
^{-1}\;\;\hbox{and}\;\;BR_{1}=1-BR_{\gamma \gamma }  \label{brslight}
\end{equation}

The probability of $N$ axions decaying in $n_{0}$\ photon pairs and $%
n_{1}=N-n_{0}$\ neutrino pairs is 
\begin{equation}
\Pr (N;n_{0})=\frac{N!}{n_{0}!(N-n_{0})!}BR_{0}^{n_{0}}(1-BR_{0})^{N-n_{0}}
\end{equation}

In the limit $N\rightarrow \infty $, by the central limit theorem, $\Pr
(N;n_{0})\rightarrow \frac{\exp \left[ -\frac{(n_{0}-N\cdot BR_{0})^{2}}{%
2NBR_{0}(1-BR_{0})}\right] }{\sqrt{2\pi N\cdot BR_{0}(1-BR_{0})}}$ and the
entropy can be written as 
\begin{eqnarray}
S_{N} &=&S_{_{0}}+\ln (2\pi N)+O(\frac{1}{N^{2}})  \label{SN} \\
S_{_{0}}(m_{A},m_{\nu },r_{\nu }) &=&\ln \left[ BR_{0}\cdot (1-BR_{0})\right]
\;.  \label{Eq:Assimptotic_entropy_one_neutrino}
\end{eqnarray}

In this case, the maximum of $S_{N}$ given by Eq.~(\ref{SN}) ocurrs for 
\begin{equation}
BR_{0}=BR_{1}=1/2\;,  \label{brs12}
\end{equation}%
in close analogy to Eq.~\ref{brssol}.

In the Appendix~\ref{app2} we derive the algebraic solutions to this
equation in details but an important difference to the previous case is
that, as shown in the Appendix~\ref{app2}, the solutions of Eq.~(\ref{brs12}%
) can only be found in terms of the ratio $z\equiv m_{\nu }/m_{A}$. This
implies that one parameter remains necessarily free in the inference method.
There are two interesting asymptotic regimes which we want to discuss. First 
$z\approx 1/2$, and second $z<<1$.

The first one is the threshold regime where $m_{A}\approx 2m_{\nu }$. This
is a type of solution which we also found in the case where the Higgs boson
has an additional decay channel to dark matter~\cite{Alves:2014ksa}.

The second interesting regime occurs when $m_{\nu }<<m_{A}$, in this case it
is possible to show that (see Appendix \ref{app2}) 
\begin{equation}
\frac{m_{\nu }}{m_{A}}\approx \frac{\sqrt{1-\sqrt{1-\frac{\alpha ^{2}}{4\pi
^{2}r_{\nu }^{2}}}}}{2}  \label{Fig:Aproximation}
\end{equation}

Moreover, as this relation should be positive, there is a lower bound on $%
r_{\nu }$ given by 
\begin{equation}
r_{\nu }=\frac{\sqrt{3\sqrt{3}}}{4\pi }\alpha =0.00132
\end{equation}%
which is a solution of the fourth order equation in $z$ explicit in Appendix %
\ref{app2}. This is compatible with the inferred couplings in the case of
three neutrinos studied in the previous section.

\begin{figure}[t]
\begin{center}
\includegraphics[width=0.5\columnwidth]{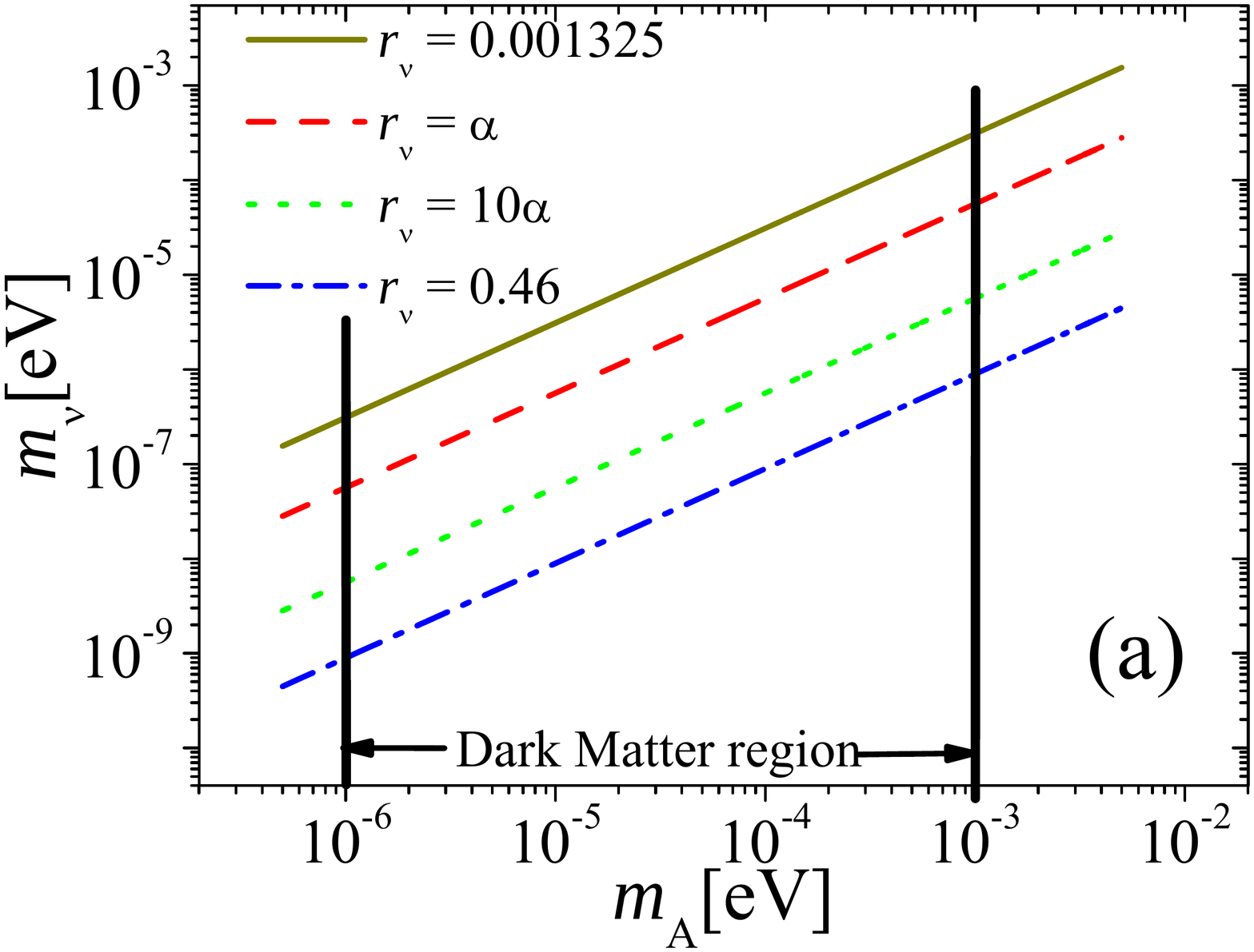}%
\includegraphics[width=0.5\columnwidth]{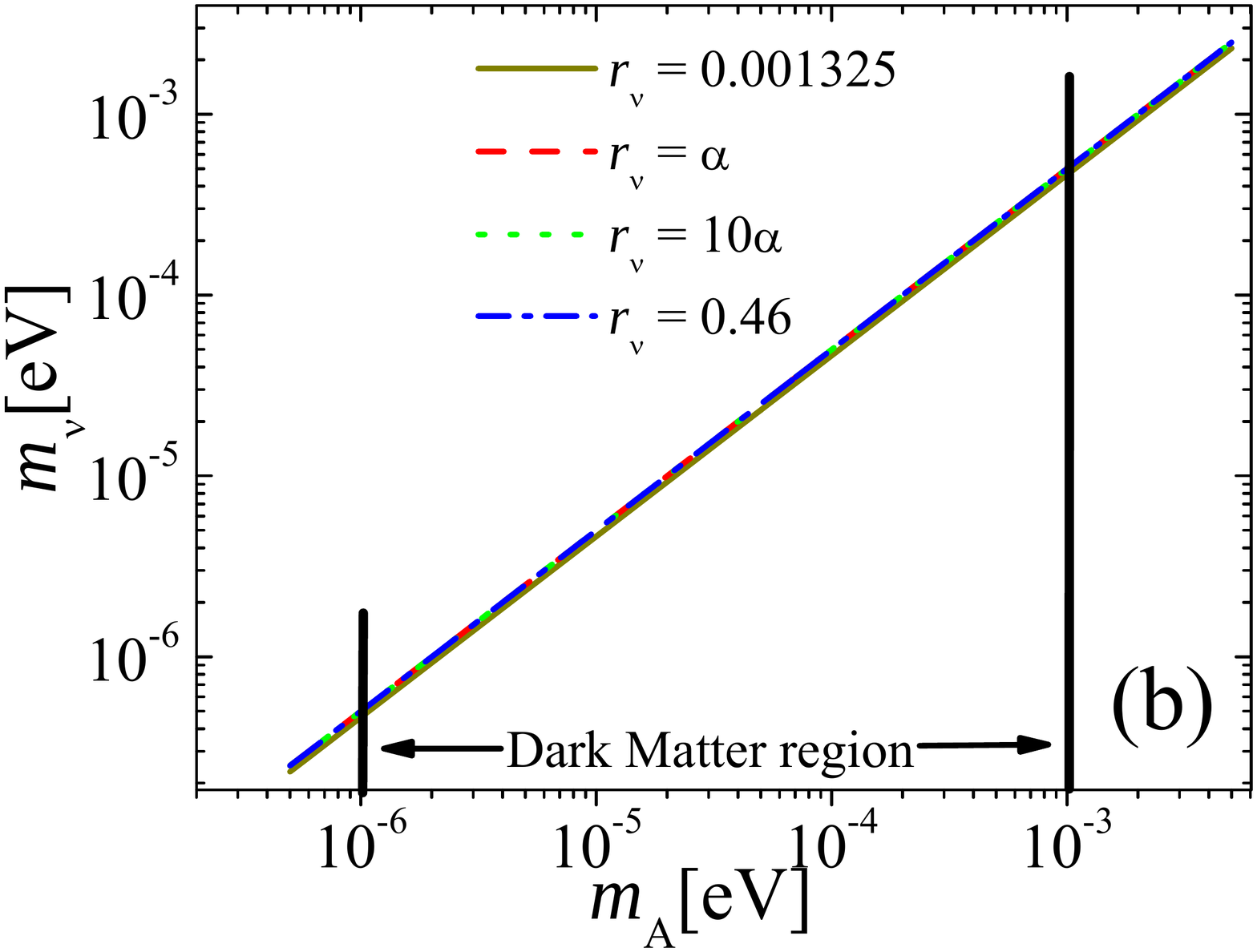}
\end{center}
\caption{Linear relations predicted by maximizing the Shannon's entropy in
the case where the axion is allowed to decay just to the lightest neutrino
and photons. In the upper plot (a) we show the $z\approx 1/2$ regime, and in
the lower plot (b), the $z<<1$ regime. }
\label{mvma}
\end{figure}

In Figure~\ref{mvma} we show the inferred masses as a function of the
axion-neutrino coupling. The upper plot in Figure \ref{mvma} displays the $%
z<<1$ regime, so the axion and the neutrino masses can be very different
depending on $r_{\nu }$ which varies from the smallest possible value of
0.00132, for which $z$ is positive, up to the maximum value of 0.46 allowed
by the axion model under consideration. In the lower plot we show the other
interesting regime where $m_{A}\approx 2m_{\nu }$, again allowing the range $%
0.00132<r_{\nu }<0.46$. In this former case, the mass inference is barely
dependent of the coupling constant. In both plots we also show the region of
the axion mass parameter where the axion can constitute, at least, part of
the cold dark matter of the Universe and be detected by a haloscope like~%
\cite{Asztalos:2011bm}, or other proposed experiments as~\cite%
{Horns:2012jf,TheMADMAXWorkingGroup:2016hpc,Millar:2016cjp}. 

Recently, the axion mass was calculated with lattice QCD methods~\cite%
{Borsanyi:2016ksw} to lie in the range $10^{-6}\ $eV$<m_{A}<1.5\times
10^{-3} $ eV\ which fits exactly in the bulk of the region shown inside the
bars in Fig.~(\ref{mvma}) . If the axion mass confirms the lattice QCD
result, one can use the MEP prediction to bound the lightest neutrino mass
and the coupling constant with the results presented in this work. Once
confirmed, this would add a strong evidence in favor of MEP as a valuable
inference tool in particle physics.

\section{Conclusions}

\label{Sec:Conclusions}

%The Maximum Entropy Principle was successfully used to obtain the most
%accurate Higgs boson mass determination to date in Ref.~\cite{Alves:2014ksa}. 

In this paper we employed MEP to a model where axions couple to photons and
neutrinos. By demanding that the Shannon entropy of an ensemble of axions
decaying to photons and neutrinos be maximized, we made inferences about the
masses of the axion taking into account its relationship with neutrinos
masses and $r_{\nu }$, which is proportional to the ratio of axion-neutrino
coupling constant and axion-photon coupling constant.

In the case where the axion decays into the three neutrino mass eigenstates,
and taking the hypothesis that the entropy is assumed to be a function of
the axion mass, the lightest neutrino mass, and the $r_{\nu }$ which is the
ratio of the axion coupling constants, under the more restrict optimization
MEP is able to make a sharp prediction: $0.1\ $ eV $\ <m_{A}<0.2\;$ eV. On
the other hand, if in the entropy $r_{\nu }$ and $m_{\nu }$ are given as
inputs, i.e., $S=S(m_{A}|r_{\nu },m_{\nu }$), considering the DFSZ model
with right-handed neutrinos, MEP jointly with astrophysical bounds constrain
the axion mass to be $0.1$~eV~$<m_{A}< 6.3$ eV.

If the axion decays only into a pair of the lightest neutrinos and photons,
the inference has two regime of solutions allowing the axion as a dark
matter candidate. First, when the axion mass is much larger than the
lightest neutrino mass, ($z<<1$), the inference of the axion mass has a
strong dependence on the ratio of the coupling constant ($r_{\nu }$) as
shown in Figure \ref{mvma} (a). On the other hand, if $m_{A}\approx 2m_{\nu
} $ ($z\approx 1/2$), the axion mass has a very weak dependence on $r_{\nu }$
as expected (see Figure \ref{mvma} (b)). For example: if $m_{A}=10^{-4}$ eV,
and $z$ $\approx 1/2$, the MEP fixes the lightest neutrino mass around $%
5\times 10^{-5}$ eV; However, if $z<<1$, the MEP predicts lightest neutrino
mass within $10^{-8}$eV$\ <m_{\nu }<10^{-5}$eV. In this case, for example,
if $r_{\nu }=\alpha $, then $m_{\nu }=7\times 10^{-6}$ eV.

Finally, we would like to stress that MEP can furnish a sharp prediction if
the neutrino mass is determined and knowing $r_{\nu }$ from some model as
shown in our Figure \ref{maximal}.

\appendix

\section{Ultraviolet model completion for to the axion-neutrinos effective
Lagrangian.}

\label{app1}

The effective Lagrangian in Eq. (\ref{axlag}) might originate from UV
completed models having a global chiral $U(1)_{PQ}$ Peccei-Quinn symmetry.
Such a symmetry is characterized by the fact it has an anomaly in the quarks
sector, leading to a mechanism for solving the strong CP problem~\cite%
{Peccei:1977hh}. The $U(1)_{PQ}$ symmetry is assumed to be spontaneously
broken at an very high energy scale giving rise to a pseudo Nambu-Goldstone
boson, the axion~\cite{Weinberg:1977ma,Wilczek:1977pj}.

A model leading to the effective Lagrangian in Eq. (\ref{axlag}) must have
neutrinos fields carrying charge of the $U(1)_{PQ}$ symmetry. In order to
have a plausible model we consider the DFSZ invisible axion model~\cite%
{Zhitnitsky:1980tq,Dine:1981rt}, and add to it three right-handed neutrinos $%
\nu_{iR}^\prime$~\footnote{%
Recently, some axion models with right-handed neutrinos have been proposed
to deal with others problems left open by the Standard Model, like for
example the neutrinos masses invoking the type-I seesaw mechanism~\cite%
{Dias:2014osa,Celis:2014zaa,Celis:2014iua,Clarke:2015bea}. In the model
example we assume here the neutrinos are taken to be of the Dirac type.
Also, we do not specify any mechanism for generating small masses for those
particles since this is not the focus of our work.}. The DFSZ model contains
a neutral singlet scalar field, $\sigma(x)$, and two Higgs doublets, $H_u(x)$%
, $H_d(x)$, with all these fields carrying charge of the $U(1)_{PQ}$
symmetry. The scalar potential constructed from these fields is assumed to
have a non-trivial minimum with the vacuum expectation values $%
\langle\sigma\rangle=v_{\sigma}/\sqrt2$, breaking the $U(1)_{PQ}$ symmetry,
and $\langle H_{u,d}\rangle=[0\,\,v_{u,d}/\sqrt2]^T$, breaking the
electroweak $SU(2)_L\otimes U(1)_Y$ symmetry. These vacuum expectation
values satisfies $\sqrt{v_u^2+v_d^2}=v=246$ GeV $\ll v_{\sigma}$. Also, it
can be seen that the decay constant is such that $f_{a^\prime}=\sqrt{%
v_{\sigma}^2+4v_u^2v_d^2/v^2}\approx v_{\sigma}$.

Let us see how the coefficient $C_{A\nu}$ of the axion-neutrinos coupling in
Eq. (\ref{axlag}) are determined in a specific model. Such a coefficient
depends on how the neutrinos fields couple to the scalar fields. Omitting
the Goldstone bosons degrees of freedom -- absorbed by the electroweak gauge
bosons -- $\sigma$ and the neutral components of $H_u$, $H_d$ can be
parametrized as 
\begin{eqnarray}
\sigma(x)&=&\frac{v_{\sigma}+\rho(x)}{\sqrt{2}}\exp\left[i\frac{a^\prime(x)}{%
f_{a^\prime}}\right], \cr H_u^{0}(x)&=&\frac{v_u+h_u(x)}{\sqrt{2}}\exp\left[%
-iX_u\frac{a^\prime(x)}{f_{a^\prime}}\right],\cr H_d^{0}(x)&=&\frac{%
v_d+h_d(x)}{\sqrt{2}}\exp\left[iX_d\frac{a^\prime(x)}{f_{a^\prime}}\right].
\label{paresc}
\end{eqnarray}
In this parametrization $h_d(x)$, $h_u(x)$, and $\rho(x)$ are CP even Higgs
fields, which are decoupled from the axion low energy effective Lagrangian
in Eq. (\ref{axlag}), and $a^\prime(x)$ the CP odd pseudo-Nambu-Goldstone
boson to be identified with the axion field $A(x)$. Such an identification
is done by mean of the axion-gluons coupling defined as 
\begin{equation}
\mathscr{L} \supset -\frac{\alpha_s}{8\pi} \,\frac{A}{f_A}\, G_{\mu\nu}^a {%
\tilde G}^{a,\mu\nu}.  \label{qcd}
\end{equation}
so that the relation in the relation 
\begin{equation}
\frac{A(x)}{f_A}=C_{{a^\prime} g}\frac{a^\prime(x)}{f_{a^\prime}}
\end{equation}
the energy scale $f_A$ is the axion decay constant, with $C_{{a^\prime} g}$
being the axion-gluon anomaly coefficient of the model. For the model we are
taking into account the axion-gluon anomaly coefficient is $C_{{a^\prime}
g}=3(X_u+X_d)=6$. The charges of the $U(1)_{PQ}$ symmetry of the scalar
fields $\sigma$, $H_u$, and $H_d$ are, respectively, $X_\sigma=1$, $-X_u=-2\,%
\mathrm{cos^2}\beta$, and $X_d=2\,\mathrm{sin^2}\beta$, with $\mathrm{tan}%
\beta=v_u/v_d$.

We assume that the axion-neutrino coupling arises from an interaction
involving the Standard Model left-handed lepton doublets, $L_i$, according
to the following Yukawa interaction 
\begin{eqnarray}
\mathscr{L}\supset F_{ij}\,\overline{L}_i \widetilde{H}_b \nu_{jR}^\prime
+h.c.  \label{yukln}
\end{eqnarray}
in which $F_{ij}$, with $i,j=1,2,3$, is a $3\times3$ matrix, and $b=u$ or $d$
if $\nu_{iR}^\prime$ couples to ${H}_u$ or ${H}_u$. We also assume that
Majorana mass terms $m_{ij}\,\overline{\nu_{iR}^{\prime c}}\nu_{jR}^{\prime
} $ are suppressed, by the $U(1)_{PQ}$ symmetry or some other extra
symmetry, relative to the Dirac mass terms arising from Eq. (\ref{yukln}).
Neutrinos masses at the sub-eV scale require small couplings $F_{ij}$ ($%
\lesssim 10^{-12}$ for $v_d\sim 100$ GeV). It is not essential to our
developments to make explicit a mechanism for achieving those small
couplings $F_{ij}$ and forbidden the Majorana mass terms, but we mention
that this could be done with discrete symmetries allowing certain
non-renormalizable operators~\cite{Dias:2005dn}. After electroweak symmetry
breakdown Eq. (\ref{yukln}) leads to 
\begin{eqnarray}
\mathscr{L}\supset F_{ij}\frac{v_b}{\sqrt{2}}\, \overline{\nu}_{iL}^\prime
\nu_{j R}^\prime \exp\left[-iX_{H_b}\frac{a^\prime(x)}{f_{a^\prime}}\right]
+h.c.  \label{yukn2}
\end{eqnarray}
With a chiral rotation $\nu_{jR}^\prime\rightarrow \nu_{j R}^\prime \exp %
\left[iX_{H_b}{a^\prime(x)}/{f_{a^\prime}}\right]$, the field $a^\prime(x)$
is removed from the above interaction leaving it as a Dirac mass term. But
the coupling of the $a^\prime(x)$ field with the neutrinos fields is induced
from the kinetic term $\overline{\nu}^\prime_{iR}i\gamma^\mu{\nu}%
^\prime_{iR}\partial_{\mu}a^\prime$ as 
\begin{equation}
\mathscr{L}\supset -\frac{X_{H_b}/C_{{a^\prime} g}}{2f_{A}} \overline{\nu}%
_{i}i\gamma^\mu\gamma_5\nu_{i}\partial_{\mu}A  \label{axnc}
\end{equation}
where ${\nu}_{i}$ denotes neutrinos mass eigenstates. Thus, defining $%
C_{A\nu}=X_{H_b}/C_{{a^\prime} g}$ in Eq. (\ref{axnc}) we have the
axion-neutrino interaction in the effective Lagrangian of Eq. (\ref{axlag}).
The coefficient $C_{\nu}$ in this model is 
\begin{eqnarray}
C_{A\nu}= \left\{%
\begin{array}{ll}
-\frac{X_u}{C_{{a^\prime} g}}=-\frac{\mathrm{cos^2}\beta}{3}, & \text{if $%
H_u $ couples to $\nu_{j R}^\prime $\,,} \\[1ex] 
\,\,\,\,\,\frac{X_d}{C_{{a^\prime} g}}=\frac{\mathrm{sin^2}\beta}{3}, & 
\text{if $H_d$ couples to $\nu_{j R}^\prime $.}%
\end{array}
\right.  \label{Can}
\end{eqnarray}

For example, if only $H_{d}$ couples to the charged right-handed charged
leptons fields the ratio of anomaly coefficients in Eq. (\ref{ctil}) is ${%
C_{a^{\prime }\gamma }}/{C_{a^{\prime }g}}=8/3$, so that ${\widetilde{C}}%
_{A\gamma }\approx 0.72$. In this case $r_{\nu }=|C_{A\nu }/{\widetilde{C}}%
_{A\gamma }|\approx 0.46\,\mathrm{cos^{2}}\beta $ (or $0.46\,\mathrm{sin^{2}}%
\beta $). On the other hand, if only $H_{u}$ couples to the charged
right-handed charged leptons fields the ratio of anomaly coefficients in Eq.
(\ref{ctil}) is ${C_{a^{\prime }\gamma }}/{C_{a^{\prime }g}}=2/3$, and ${%
\widetilde{C}}_{A\gamma }\approx -1.28$. In this case $|r_{\nu }|=|C_{A\nu }/%
{\widetilde{C}}_{A\gamma }|\approx 0.26\,\mathrm{cos^{2}}\beta $ (or $0.26\,%
\mathrm{sin^{2}}\beta $). The expressions for the coefficients ${%
C_{a^{\prime }\gamma }}/{C_{a^{\prime }g}}$ can be found in~\cite%
{Dias:2014osa}.

Just for completion we present the axion-electron coupling in the DFSZ model
we are considering. Proceeding with a chiral rotation in the right-handed
charged leptons singlet fields $e_{iR}^\prime\rightarrow e_{i R}^\prime \exp %
\left[-iX_{H_b}{a^\prime(x)}/{f_{a^\prime}}\right]$ the coefficients of the
axion-electron derivative coupling is 
\begin{eqnarray}
C_{Ae}= \left\{%
\begin{array}{ll}
\,\,\,\,\,\frac{X_u}{C_{{a^\prime} g}}=\frac{\mathrm{cos^2}\beta}{3}, & 
\text{if $H_u$ couples to $e_{iR}^\prime $\,,} \\[1ex] 
-\frac{X_d}{C_{{a^\prime} g}}=-\frac{\mathrm{sin^2}\beta}{3}, & \text{if $%
H_d $ couples to $e_{iR}^\prime $.}%
\end{array}
\right.  \label{eac}
\end{eqnarray}

\section{Algebraic solutions in the one neutrino case}

\label{app2}

The solution of this equation $BR_{\gamma \gamma }(C_{\nu },m_{A},m_{\nu
})=1/2$\ leads to equation 
\begin{equation}
G(z)=z^{2}(1-4z^{2})^{1/2}=\alpha ^{2}/(32\pi ^{2}r_{\nu }^{2})
\label{Eq:Main_z}
\end{equation}%
where $z^{2}=\frac{m_{\nu }^{2}}{m_{A}^{2}}$. Writing such equation in $%
x=z^{2}$, a simple cubic equation appears:

\begin{equation}
ax^{3}+bx^{2}+cx+d(r_{\nu })=0  \label{Eq: Cubic}
\end{equation}%
where $a=4$, $b=-1$, $c=0$ and, $d(r_{\nu })=\left( 32\pi ^{2}\frac{r_{\nu
}^{2}}{\alpha ^{2}}\right) ^{-2}$.

Denoting $p=\frac{c}{a}-\frac{b^{2}}{3a^{2}}=-\frac{1}{48}$, and $q(%
\overline{C_{\nu }})=\frac{2b^{3}}{27a^{3}}-\frac{bc}{3a^{2}}+\frac{d(r_{\nu
})}{a}=\frac{-2}{27\cdot 4^{3}}+\frac{1}{4}\left( 32\pi ^{2}\frac{r_{\nu
}^{2}}{\alpha ^{2}}\right) ^{-2}=\allowbreak \frac{1}{4096\pi ^{4}}\frac{%
\alpha ^{4}}{r_{\nu }^{4}}-\frac{1}{864}$ and defining 
\begin{equation}
\begin{array}{lll}
\Delta (r_{\nu }) & = & q(r_{\nu })^{2}+\frac{4p^{3}}{27} \\ 
&  &  \\ 
& = & \left( \frac{1}{4096\pi ^{4}}\frac{\alpha ^{4}}{r_{\nu }^{4}}-\frac{1}{%
864}\right) ^{2}-\frac{1}{746\,496}%
\end{array}
\label{Eq:Delta}
\end{equation}%
and by solving this cubic equation we have 3 solutions:%
\begin{equation}
x_{1}(r_{\nu })=t(r_{\nu })+\frac{1}{12}  \label{Eq:x1}
\end{equation}%
where $t(r_{\nu })=\left( -\frac{q(r_{\nu })}{2}+\frac{1}{2}\sqrt{\Delta
(r_{\nu })}\right) ^{1/3}+\left( -\frac{q(r_{\nu })}{2}-\frac{1}{2}\sqrt{%
\Delta (r_{\nu })}\right) ^{1/3}$,

\begin{equation}
x_{2}(r_{\nu })=-\frac{t(r_{\nu })}{2}+\sqrt{\frac{t(r_{\nu })^{2}}{4}+\frac{%
q(r_{\nu })}{t(r_{\nu })}}+\frac{1}{12}  \label{Eq:x2}
\end{equation}%
and

\begin{equation}
x_{3}(r_{\nu })=-\frac{t(r_{\nu })}{2}-\sqrt{\frac{t(r_{\nu })^{2}}{4}+\frac{%
q(r_{\nu })}{t(r_{\nu })}}+\frac{1}{12}  \label{Eq:x3}
\end{equation}

\begin{figure}[th]
\begin{center}
\includegraphics[width=0.4\textwidth,height=0.3\textheight]{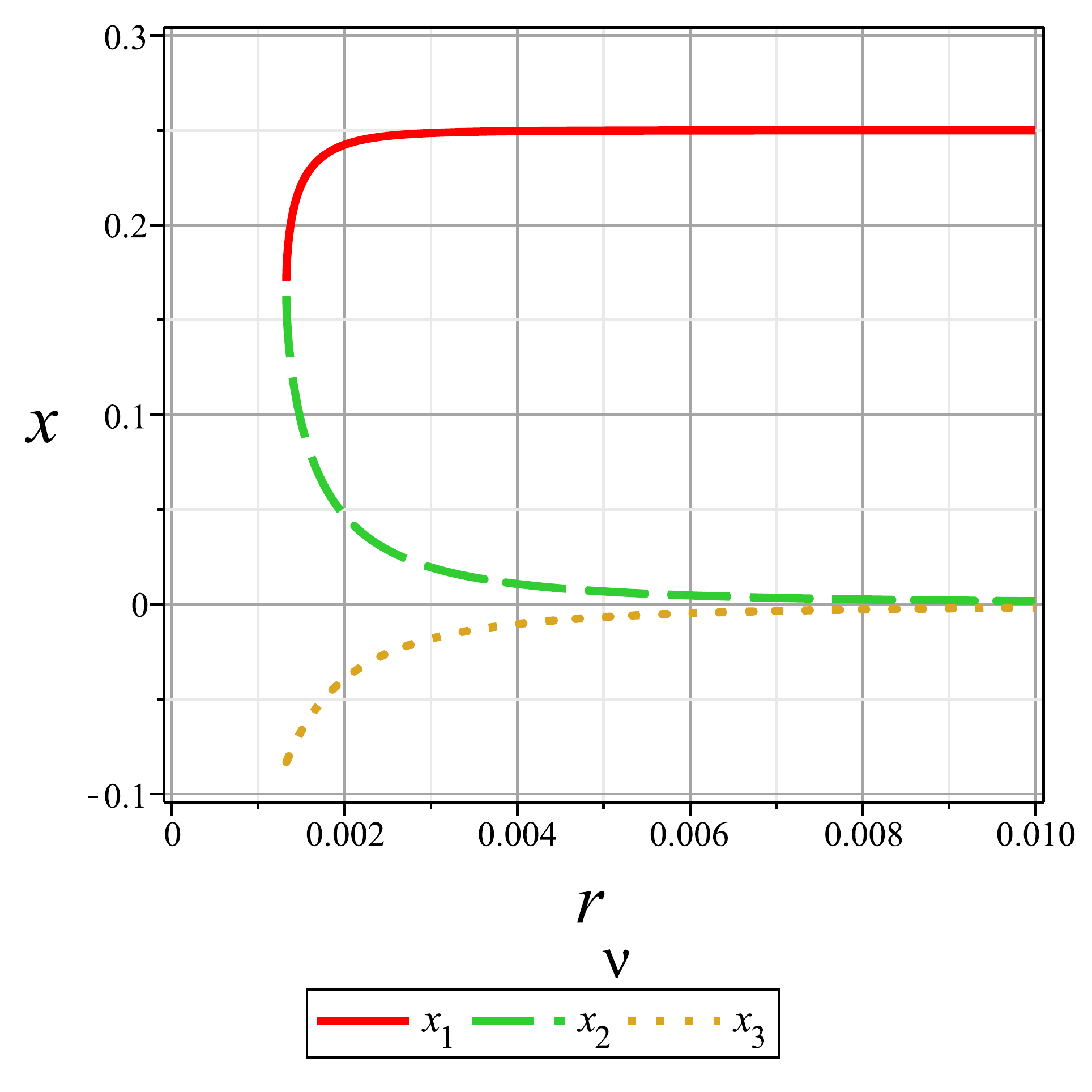}
\end{center}
\caption{Three roots of Eq. \protect\ref{Eq: Cubic} for $x=z^{2}=\left( m_{%
\protect\nu }/m_{A}\right) ^{2}$. }
\label{Fig:roots}
\end{figure}

We must observe that only $x_{1}$ and $x_{2}$ are positive numbers while $%
x_{3}$ is negative, as we observe in Figure \ref{Fig:roots} that show $x_{1}$%
, $x_{2}$, and $x_{3}$ as function of $r_{\nu }$.

So, both solutions $x_{1}$ and $x_{2}$ determine two direct relations
between the axion mass and the neutrino mass: $m_{A}=x_{1}^{-1/2}(r_{\nu
})m_{\nu }$ and $m_{A}=x_{2}^{-1/2}(r_{\nu })m_{\nu }$. By Figure \ref%
{Fig:roots}, we can obtain the two asymptotic cases by considering two
situations:

\textbf{Situation I}: $z<<1$;

For such situation, we can consider the approximation: $G(z)=z^{2}\sqrt{%
1-4z^{2}}\approx z^{2}-2z^{4}=\frac{\alpha ^{2}}{32\pi ^{2}r_{\nu }^{2}}$,
which leads to a simple relation:

\begin{equation}
\frac{m_{\nu }}{m_{A}}\approx \frac{\sqrt{1-\sqrt{1-\frac{\alpha ^{2}}{4\pi
^{2}r_{\nu }^{2}}}}}{2}  \label{Fig:Aproximation}
\end{equation}%
valid for$\ r_{\nu }\geq \frac{\sqrt{3\sqrt{3}}\alpha }{4\pi }\simeq 0.00132$%
, which asymptotically behaves as 
\begin{equation}
\frac{m_{\nu }}{m_{A}}\sim \frac{1}{4\sqrt{2}}\frac{\alpha }{\pi r_{\nu }}
\label{Eq:MnuvsMA}
\end{equation}%
and therefore $\frac{m_{\nu }}{m_{A}}\rightarrow 0$ when $r_{\nu
}\rightarrow \infty $.

\textbf{Situation II}: $z\approx 1/2$;

In this case \ we consider the approximation: $G(z)=z^{2}\sqrt{1-4z^{2}}%
=z^{2}\sqrt{(1-2z)(1+2z)}\approx \frac{1}{4}\sqrt{2(1-2z)}=\frac{\alpha ^{2}%
}{32\pi ^{2}r_{\nu }^{2}}$. We have 
\begin{equation}
\frac{m_{\nu }}{m_{A}}\sim \frac{1}{2}-\frac{\alpha ^{4}}{256\pi ^{4}r_{\nu
}^{4}}\rightarrow \frac{1}{2}  \label{Eq:MnuvsMA_II}
\end{equation}%
for higher $r_{\nu }$.

\begin{figure*}[th]
\begin{center}
\includegraphics[width=0.4\textwidth,height=0.3%
\textheight]{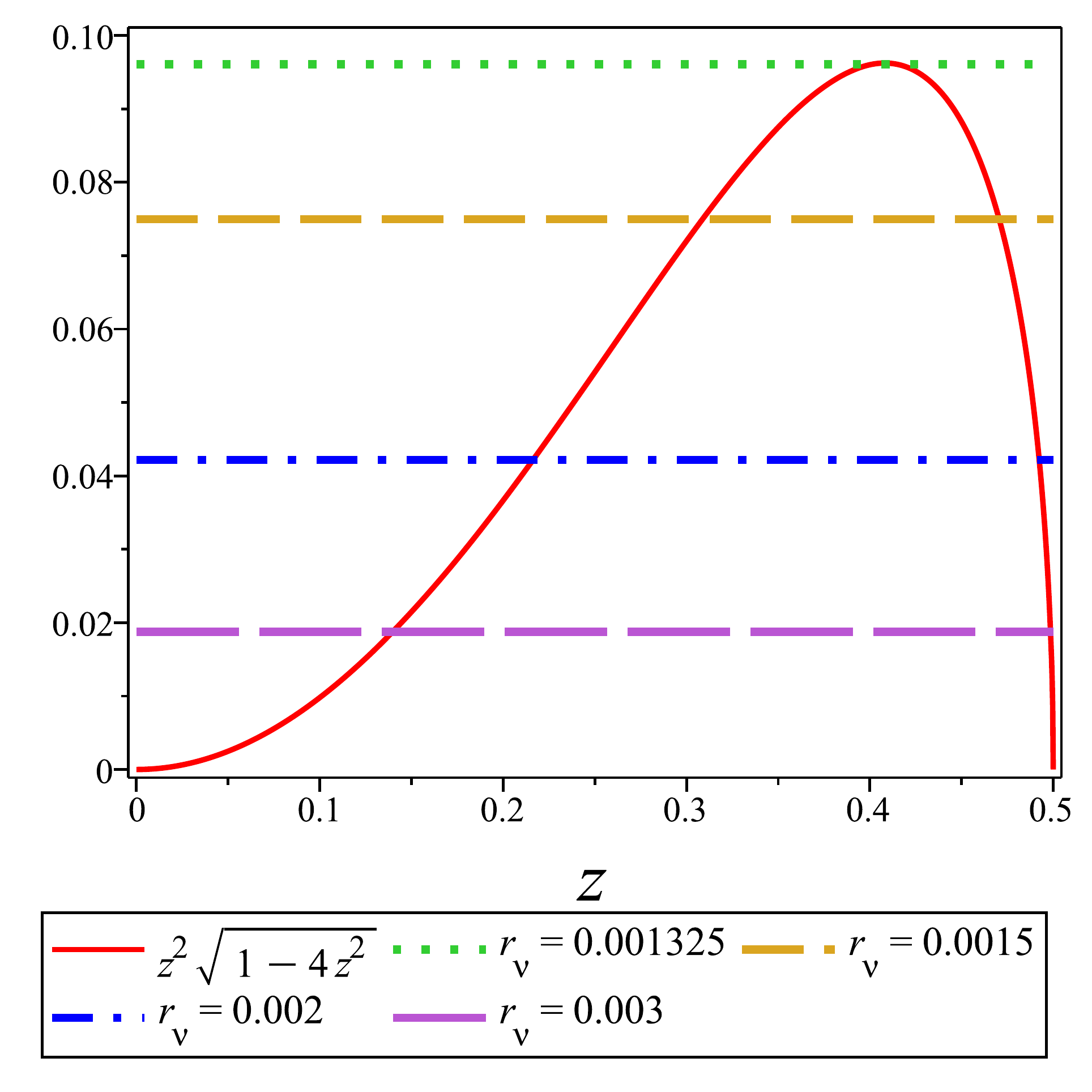}
\includegraphics[width=0.4%
\textwidth,height=0.3\textheight]{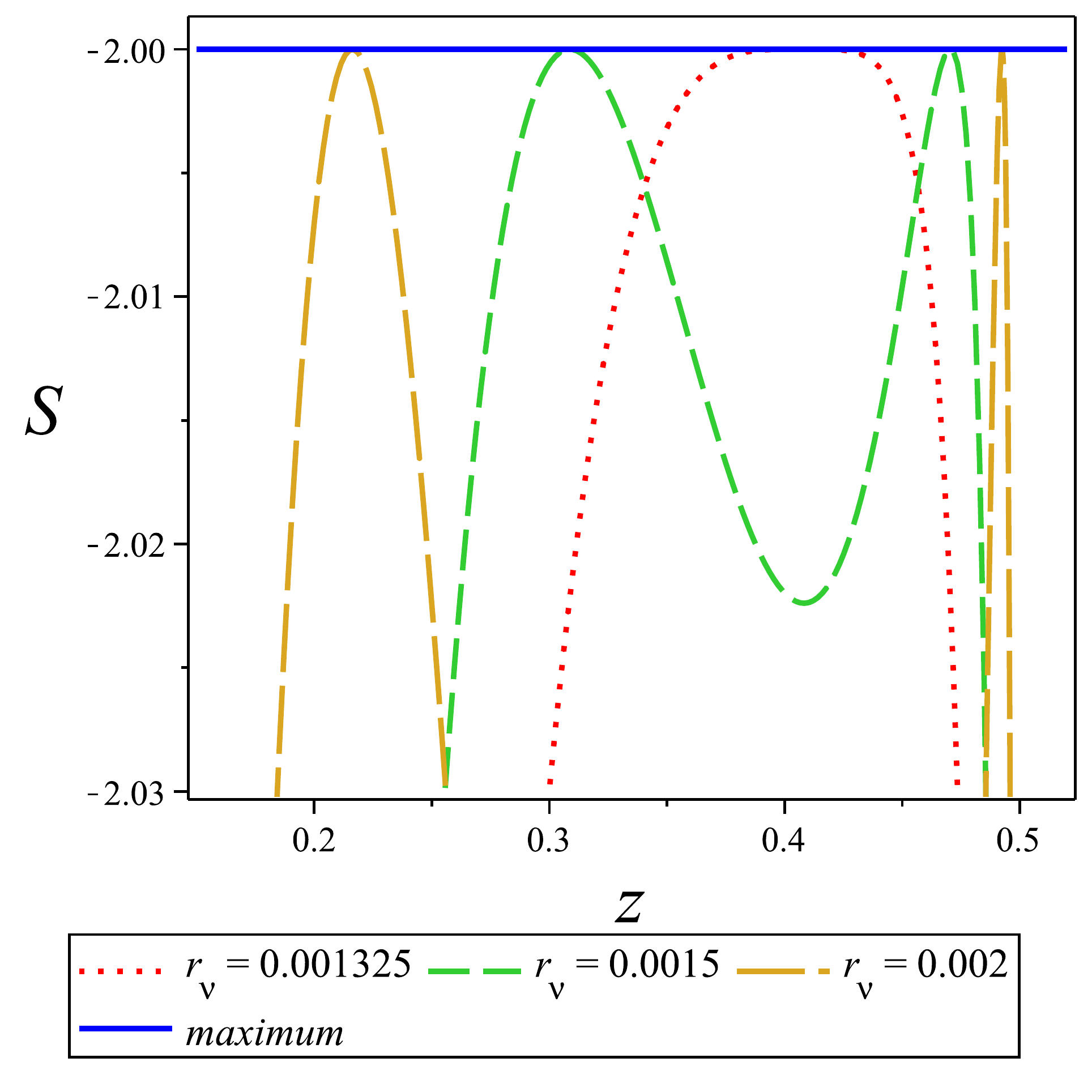}
\end{center}
\caption{(Left Plot) This plot shows the condition to maximum: $BR_{0}=1/2$.
The parallel lines represent different values of $\left( 32\protect\pi ^{2}%
\frac{r_{\protect\nu }^{2}}{\protect\alpha ^{2}}\right) ^{-1}$while the
curve corresponds tot he plot of $z^{2}(1-4z^{2})$. We can see that given $%
r_{\protect\nu }$ we obtain two distinct values of $z$. (Right plot)$\ \ $%
This plot corresponds to the same situation of the left-top plot looking for
the entropy. The maximum $S_{0}=-2$ occurs for two different values of $z$
corresponding to the intersection points in the previous figure. It is
important to see that $r_{\protect\nu }=0.00132$ corresponds to an unique $%
z$ value which is exactly $1/\protect\sqrt{6}\simeq 0.408$.}
\label{Fig:set_of_one_neutrino_plots}
\end{figure*}

Let us better explore some important points. In Figure \ref%
{Fig:set_of_one_neutrino_plots} we can observe such result from two
different ways. Left plot shows the maximum condition $BR_{0}=1/2$. The
parallel lines represent different values of $\left( 32\pi ^{2}\frac{r_{\nu
}^{2}}{\alpha ^{2}}\right) ^{-1}$while the curve corresponds to the plot of $%
z^{2}(1-4z^{2})^{1/2}$. We can see that given $r_{\nu }$ we obtain two
distinct values of $z$ (two intersections). For $r_{\nu }\simeq 0.00132$ we
have a only intersection point which corresponds to the $z=1/\sqrt{6}\simeq
0.408\,$. In Right plot we check the same situation but looking for the
entropy. We consider $S_{0}$ in the base two and not $e$. This implicates
that maximum of the $S_{0}$ is $-2$ ($BR_{0}=BR_{1}=1/2$). This global
maximum occurs for two distinct $z$-values for different values of $r_{\nu }$
except by $z=$ $1/\sqrt{6}$ (or $r_{\nu }=0.00132$).

\end{document}